\documentclass[11pt,letterpaper]{article}
\usepackage[T1]{fontenc}
\usepackage[utf8]{inputenc}
\usepackage{lmodern}
\usepackage[margin=1in]{geometry}
\usepackage{amsmath,amssymb,amsthm}
\usepackage{microtype}
\usepackage{graphicx,tikz}
\usepackage{float,enumitem,needspace}
\usepackage{hyperref}
\hypersetup{
  colorlinks=true,linkcolor=black,citecolor=black,urlcolor=black,
  pdftitle={A Polynomial-Time Algorithm for ARRIVAL},
  pdfsubject={Periodic routing, exact fixed points, and random access},
  pdfauthor={Chenglin Fan, Jeonseong Kim}
}
\newtheorem{theorem}{Theorem}[section]
\newtheorem{lemma}[theorem]{Lemma}
\newtheorem{corollary}[theorem]{Corollary}
\numberwithin{equation}{section}

\newcommand{\ARRIVAL}{\textnormal{\upshape\scshape Arrival}}

\newcommand{\e}{\mathbf e}
\newcommand{\norm}[1]{\left\lVert #1\right\rVert_1}

\floatstyle{ruled}
\newfloat{algorithm}{tbp}{loa}
\floatname{algorithm}{Algorithm}
\newcounter{algline}
\newcounter{algdepth}
\newlength{\algindent}
\newenvironment{algorithmic}[1][1]{%
  \small\setcounter{algdepth}{0}%
  \begin{list}{\scriptsize\arabic{algline}:}{%
    \usecounter{algline}\setlength{\leftmargin}{1.9em}%
    \setlength{\labelwidth}{1.35em}\setlength{\labelsep}{0.55em}%
    \setlength{\itemsep}{1.5pt}\setlength{\parsep}{0pt}%
    \setlength{\topsep}{3pt}\setlength{\partopsep}{0pt}}%
}{\end{list}}
\newcommand{\State}{\item\hspace*{\dimexpr\value{algdepth}\algindent\relax}}

\newcommand{\Require}{\item[]\textbf{Input:} }
\newcommand{\Ensure}{\item[]\textbf{Output:} }
\newcommand{\Call}[2]{\text{\normalfont\textsc{#1}}(#2)}
\newcommand{\Return}{\textbf{return}\ }
\newcommand{\While}[1]{\State\textbf{while} #1 \textbf{do}\stepcounter{algdepth}}
\newcommand{\EndWhile}{\addtocounter{algdepth}{-1}\State\textbf{end while}}
\newcommand{\ForAll}[1]{\State\textbf{for each} #1 \textbf{do}\stepcounter{algdepth}}

\newcommand{\EndFor}{\addtocounter{algdepth}{-1}\State\textbf{end for}}
\newcommand{\If}[1]{\State\textbf{if} #1 \textbf{then}\stepcounter{algdepth}}

\newcommand{\EndIf}{\addtocounter{algdepth}{-1}\State\textbf{end if}}
\newcommand{\Loop}{\State\textbf{loop}\stepcounter{algdepth}}
\newcommand{\EndLoop}{\addtocounter{algdepth}{-1}\State\textbf{end loop}}

\begin{document}

\begingroup
\renewcommand{\thefootnote}{}
{\centering
  {\LARGE\bfseries A Polynomial-Time Algorithm for ARRIVAL\par}
  \vspace{1.0em}
  {\large Chenglin Fan\qquad Jeonseong Kim\footnotemark\par}
}
\footnotetext{Authors are listed in alphabetical order.}
\endgroup

\vspace{0.3em}

\begin{abstract}
In ARRIVAL, a token moves through a directed graph from a specified starting vertex, alternating between the two outgoing arcs of each vertex on successive departures. The problem asks whether the token ever reaches a given destination. Although the walk can be exponentially long, we give a deterministic polynomial-time algorithm, resolving an open problem. We first average the local routing rules and then restore the original rules on selected periods. The number of refinement rounds is at most the number of nonterminal vertices. The algorithm extends to explicitly given cyclic successor lists. It also computes exact departure counts when every nonterminal vertex has a path to an absorbing terminal and recovers the configuration of the walk at any binary-encoded time.
\end{abstract}

\section{Introduction}\label{sec:intro}

Consider a token moving through a directed graph with two ordered outgoing arcs at each nonterminal vertex. On its first departure from a vertex it takes the first arc, on its second departure the second arc, and thereafter alternates between the two. The question in \ARRIVAL\ is whether the token, starting at a specified origin, ever reaches a given destination. The problem was introduced by Dohrau, G\"artner, Kohler, Matou\v{s}ek, and Welzl~\cite{dohrau2017}, who proved that it belongs to $\mathsf{NP}\cap\mathsf{coNP}$. It is an instance of a rotor-router walk, a deterministic analogue of a random walk~\cite{holroyd2008}. The difficulty of predicting the walk is already apparent on a chain of $n$ switches: let the first arc at each switch return to the origin, and the second advance to the next switch or, at the end of the chain, to a terminal. The token leaves the last switch twice, its predecessor four times, and the origin $2^n$ times. The departure counts have short binary representations, even though computing them by simulation takes exponential time.

We consider a single token in the explicit zero-player periodic routing model of Fearnley et al.~\cite{fearnley2021}. Let $V$ be the set of $n\ge1$ internal vertices, let $S$ be the set of absorbing terminals, and let $o\in V$ be the origin. At each vertex $v\in V$, a cyclic list of $d_v\ge1$ successors in $V\cup S$ specifies the order of departures. Repetitions are allowed, and the initial phase of each list is part of the input. Write $D=\sum_{v\in V}d_v$ for the total length of the lists. A network is \emph{stopping} if every internal vertex has a directed path of listed arcs to a terminal. We prove the following result.

\begin{theorem}\label{thm:main}
Designated-destination reachability in an explicit periodic routing network with $n$ internal vertices and total successor-list length $D$ is decidable in deterministic polynomial time. The algorithm uses $O(n^4D)$ exact rational arithmetic operations and has polynomial bit complexity. On a stopping network it also returns the exact departure count at each internal vertex. In particular, $\ARRIVAL\in\mathsf P$, with an $O(n^5)$ arithmetic bound for binary switches.
\end{theorem}

The bit complexity is polynomial in the explicit input length. We also use departure counts to recover the configuration of the original walk after an arbitrary binary-encoded number of moves (Theorem~\ref{thm:random}).

\subsection{Complexity and previous algorithms}\label{sec:prior}

Early work on \ARRIVAL\ focused on its complexity. Karthik~\cite{karthik2017} placed a search version of the problem in $\mathsf{PLS}$. G\"artner et al.~\cite{gartner2018} subsequently placed the decision problem in $\mathsf{UP}\cap\mathsf{coUP}$ and the search version in $\mathsf{CLS}$, and gave a randomized algorithm with expected running time $O(1.4143^n)$. Fearnley et al.~\cite{fearnley2021} studied more general switching games, distinguishing the zero-player problem from the substantially harder variants with one or two players.

The first general subexponential algorithm, by G\"artner, Haslebacher, and Hoang~\cite{gartner2021}, runs in time $n^{O(\sqrt n)}$. The idea is to turn a selected set of vertices into terminals, guess their departure counts, and solve the resulting multi-token routing instance. A guess is valid if the number of tokens arriving at each selected vertex equals the guessed number of departures there. Finding consistent guesses leads to a multidimensional fixed-point search on the selected vertices. Haslebacher~\cite{haslebacher2025} gave a recursive formulation in which the count of one vertex is determined by binary search, and used it to derive, among other bounds, a $2^{O(k\log^2 n)}$-time algorithm for graphs of treewidth $k$. Ghorbani, Lopes, and Mnich~\cite{ghorbani2026} subsequently obtained time $2^{O(n^{1/3}\log^{2/3}n)}$ by guessing vertices in blocks and approximating centerpoints to organize the search in each block. In these algorithms, evaluating the consistency map requires solving another routing instance. The block method trades recursion depth for the cost of searching within each block, yielding the cube-root bound.

Other algorithms exploit the structure of particular graph families. Auger, Coucheney, and Duhaz\'e~\cite{auger2022} gave almost-linear-time algorithms for tree-like multigraphs, based on return flows in subtrees. For the multi-token generalization, Auger et al.~\cite{auger2023} gave polynomial-time algorithms on path multigraphs with a prescribed rotor order. Ghorbani, Leander Hoff, and Mnich~\cite{ghorbani2025} extended this direction to quasi-polynomial time on tree-like multigraphs with arbitrary rotor orders and polynomial time in several special cases. Their tree-like algorithm also uses affine approximations of return-flow functions. More recently, G\"artner, Haslebacher, and Hoang~\cite{ladders2026} obtained polynomial time for ladders, a family on which earlier general algorithms may take their worst-case running time. Our algorithm handles arbitrary directed graphs in the explicit single-token model.

\subsection{The contraction approach}\label{sec:contraction-prior}

Haslebacher~\cite{haslebacher2025} reduced \ARRIVAL\ and its multi-token generalization to approximating a fixed point of a continuous $\ell_1$-contraction. The reduction replaces integer departure counts by continuous ones, uses a discount factor just below one to ensure uniqueness of the fixed point, and recovers the terminal arrival counts from a sufficiently accurate approximation. This formulation works directly on the original graph. The discount may be exponentially close to one, however, so the convergence bound for ordinary fixed-point iteration is exponential.

Chen, Li, and Yannakakis~\cite{chen2024} gave a polynomial query bound for $\ell_\infty$-contractions. Haslebacher, Lill, Schnider, and Weber~\cite{haslebacher2025queries} extended this to all $\ell_p$-norms, $1\le p\le\infty$: for a map on the $m$-dimensional unit cube with contraction factor $\lambda<1$, they find a point of residual at most $\varepsilon$ using
\[
 O\!\left(m^2\left(\log\frac1\varepsilon+\log\frac1{1-\lambda}\right)\right)
\]
queries, covering the parameter range of the \ARRIVAL\ reduction. The query bound controls the number of evaluations; computing generalized centerpoints remains a bottleneck for polynomial running time. Batziou et al.~\cite{batziou2025} also exploit monotonicity to obtain algorithms for $\ell_\infty$-contractions with efficient individual queries and dimension-dependent complexity. Routing gives rise to an $\ell_1$-contraction, for which the computation of a fixed point remains the central issue in this approach.

We retain the discounted formulation and exploit the local routing functions directly. These functions can have a breakpoint at every integer departure count. In the chain above, for example, Lemma~\ref{lem:rounding} places the final discounted count at the origin within one of $2^n$, so a direct continuation would traverse exponentially many pieces. Our algorithm represents only selected periods of the exact routing functions and uses their averages elsewhere. The exact and averaged outflows agree coordinatewise at each period boundary, which allows us to bound the number of periods that must be restored.

\subsection{Technical overview}\label{sec:overview}

\begin{figure}[t]
\centering
\begin{tikzpicture}[x=0.84cm,y=0.84cm,font=\small]
  \foreach \k in {0,1,2}{\fill[black!5] (\k,\k) rectangle ({\k+1},{\k+1});}
  \draw[->] (-0.1,0)--(3.6,0) node[right] {to $w_0$};
  \draw[->] (0,-0.1)--(0,3.6) node[above] {to $w_1$};
  \draw[densely dashed,thick,black!55] (0,0)--(3,3);
  \draw[very thick] (0,0)--(1,0)--(1,1)--(2,1)--(2,2)--(3,2)--(3,3);
  \foreach \k in {1,2,3}{
    \fill (\k,\k) circle (1.6pt);
    \draw (\k,0.05)--(\k,-0.05) node[below] {\(\k\)};
    \draw (0.05,\k)--(-0.05,\k) node[left] {\(\k\)};
  }
  \draw[very thick] (4.7,2.7)--(5.25,2.7);
  \node[anchor=west] at (5.4,2.7) {$r_v$ (exact)};
  \draw[densely dashed,thick,black!55] (4.7,2.05)--(5.25,2.05);
  \node[anchor=west] at (5.4,2.05) {$h_v$ (averaged)};
  \fill (4.97,1.4) circle (1.6pt);
  \node[anchor=west] at (5.4,1.4) {Common period endpoint};
\end{tikzpicture}
\caption{Local outflows of a switch with successor list $(w_0,w_1)$,
where $w_0\ne w_1$. The axes are the amounts sent to the two successors
(for either rule). Each shaded square contains the outflows of one
period. Points in different squares are coordinatewise comparable.}
\label{fig:periods}
\end{figure}
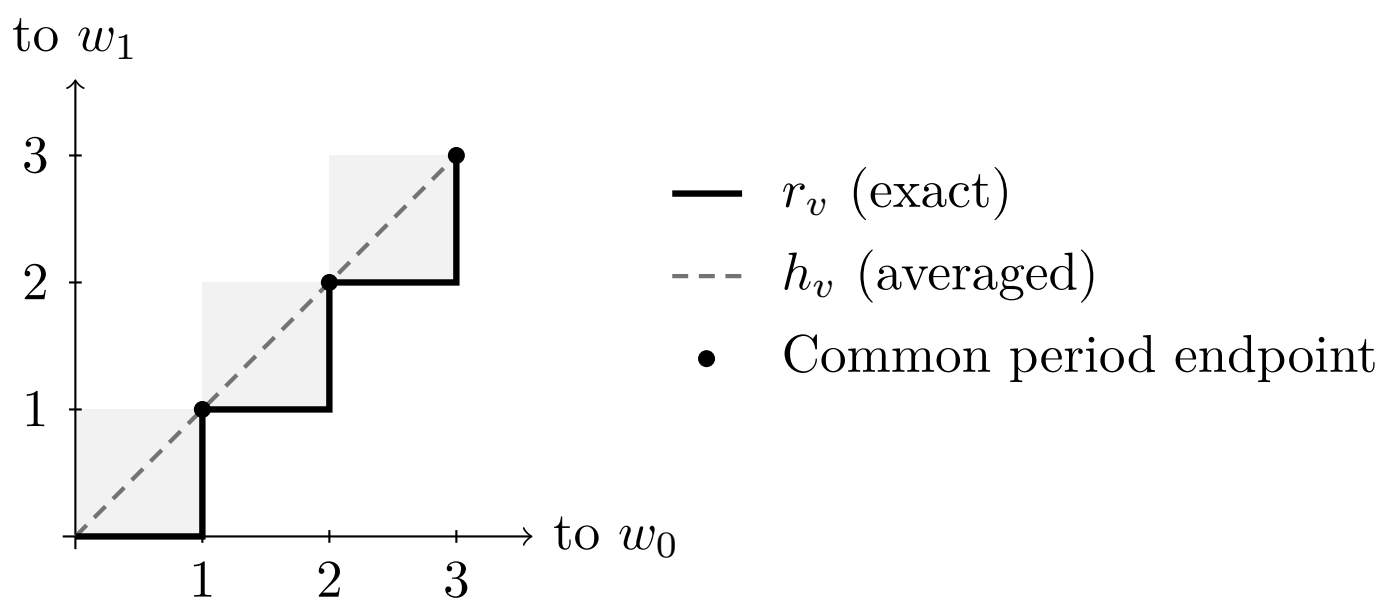

Let $r_v(z)$ be the vector of amounts sent to the successors of $v$ after $z\ge0$ departures, extended linearly between integer values, and let $h_v(z)$ split the same amount in proportion to the frequencies of the successors in one period. Both vectors include the coordinates of absorbing terminals. Their entries sum to $z$, and both rules are coordinatewise nondecreasing. The two rules agree at the period boundaries $kd_v$ for integers $k\ge0$. For example, at a binary switch with distinct successors the exact outflow after seven departures is $(4,3)$, while the averaged outflow is $(3.5,3.5)$. After eight departures both are $(4,4)$. Consequently, whenever a period boundary lies between two input values, the outflow at the smaller input is coordinatewise at most the outflow at the larger input, regardless of whether either is evaluated with the exact or averaged rule. Figure~\ref{fig:periods} illustrates this observation.

\emph{Period refinement.} Write $H$ for the map that sums the exact local outflows and keeps the internal inflow coordinates, $\e_o$ for the unit vector at the origin, and $q\in(0,1)$ for the rational discount. The exact discounted departure vector is the unique solution $x=\e_o+qH(x)$. Initially we replace every local rule by its average; subsequently we restore the exact rule on a finite set $W_v$ of periods at vertex $v$, leaving the average everywhere else. Write $W=(W_v)_{v\in V}$ for these sets, $r_v^W$ for the hybrid local rule, $H_W$ for the corresponding inflow map, and $y=\e_o+qH_W(y)$ for the hybrid solution. If at every vertex the two full outflow vectors $r_v(x_v)$ and $r_v^W(y_v)$ are coordinatewise comparable, their $\ell_1$ distance is simply $|x_v-y_v|$, since their coordinate sums are $x_v$ and $y_v$. Writing $\norm{\cdot}$ for the $\ell_1$ norm, the fixed-point equations and the triangle inequality then give
\[
 \norm{x-y}\le q\sum_{v\in V}\norm{r_v(x_v)-r_v^W(y_v)}=q\norm{x-y}.
\]
As $q<1$, this is possible only if $x=y$. If $y\ne x$, some vertex therefore has incomparable outflows. The common boundaries imply that $x_v$ and $y_v$ lie strictly inside the same period at that vertex. This period is outside $W_v$, since the two rules coincide on restored periods. We therefore restore the period containing $y_v$ at every vertex. At least one of the newly restored periods contains the corresponding coordinate of $x$. Once such a period is restored, the outflows at that vertex are comparable in every subsequent round: a later hybrid value either belongs to the restored period or is separated from $x_v$ by a common boundary. Each unsuccessful test thus restores a period containing a true coordinate at a vertex where this has yet to happen. The restored periods are retained, so at most $n$ tests can fail, regardless of the numerical error of the hybrid solution.

\emph{Solving the hybrid system.} Between restored periods the averaged rule is a single affine function, irrespective of the length of the intervening interval. To solve a hybrid system, increase the injection from zero to one and let $y(\alpha)$ be the solution at injection $\alpha\e_o$ for $\alpha\in[0,1]$. Monotonicity implies that each coordinate of $y(\alpha)$ is nondecreasing. While the coordinates remain in their current local pieces, the equation is a linear system with a nonsingular matrix; its solution is affine in $\alpha$, so the next breakpoint can be computed directly. Each event advances at least one coordinate to a new piece, and monotonicity limits the total number of events to the number of represented breakpoints, which is $O(nD)$ per round. Solving a linear system at each event gives an $O(n^6)$ arithmetic bound for binary switches. When a single coordinate reaches an endpoint, just one matrix column changes. The Sherman--Morrison identity~\cite{sherman1950} reduces the cost of this change from cubic to quadratic in $n$. The resulting exact discounted fixed point can be rounded to the integer departure counts using a discount of polynomial bit length.

Section~\ref{sec:model} gives the continuous formulation and the rounding lemma. Section~\ref{sec:algorithm} presents the algorithm and its complexity analysis. Section~\ref{sec:random} uses departure counts to recover the configuration at a specified time.

\section{Periodic routing and discounted flow}\label{sec:model}

\subsection{Continuous routing maps}\label{sec:maps}

Let the cyclic successor list at $v\in V$ be
$\sigma_v(0),\ldots,\sigma_v(d_v-1)\in V\cup S$. Repeated successors and
self-loops are allowed. By rotating each list to its initial phase, we may
assume the first departure from $v$ uses $\sigma_v(0)$. The case of an absorbing origin is immediate. For vectors,
inequalities are coordinatewise, and $\norm{\cdot}$ denotes the
$\ell_1$ norm. We use $\e_w$ for the unit vector at coordinate $w$ in the
ambient coordinate space indicated by the expression. In particular,
local outflows lie in $\mathbb R^{V\cup S}$, while internal inflows and
vectors of departure counts lie in $\mathbb R^V$. Let $\pi$ denote the
projection from $\mathbb R^{V\cup S}$ onto the internal coordinates.

For each $v\in V$, define its multiplicity vector $m_v\in\mathbb Z_{\ge0}^{V\cup S}$
by letting $(m_v)_w$ count occurrences of $w$ in the list at $v$. For an
integer $j\ge0$, the exact outflow is
\[
 r_v(j)=\sum_{i=0}^{j-1}\e_{\sigma_v(i\bmod d_v)},
\]
where an empty sum is zero. Extend $r_v$ to all real $z\ge0$ by linear
interpolation between consecutive integers. The averaged outflow is
$h_v(z)=(z/d_v)m_v$. Both maps are coordinatewise nondecreasing, have
coordinate sum $z$ at input $z$, and satisfy
\begin{equation}\label{eq:boundary}
 r_v(kd_v)=h_v(kd_v)=km_v\qquad(k\in\mathbb Z_{\ge0}).
\end{equation}
In particular, both rules lie in the coordinatewise interval
$[km_v,(k+1)m_v]$ throughout the $k$th period
$[kd_v,(k+1)d_v]$.

Let $W_v\subseteq\mathbb Z_{\ge0}$ be a finite set of restored period
indices and write $W=(W_v)_{v\in V}$. The hybrid rule $r_v^W$ equals $r_v$
on the closed periods with indices in $W_v$ and $h_v$ elsewhere. By
\eqref{eq:boundary}, this definition is unambiguous, and $r_v^W$ is
continuous, coordinatewise nondecreasing, and has coordinate sum $z$ at
input $z$. Define the exact and hybrid internal inflow maps on
$\mathbb R_{\ge0}^V$ by
\[
 H(z)=\pi\sum_{v\in V} r_v(z_v),\qquad
 H_W(z)=\pi\sum_{v\in V} r_v^W(z_v).
\]

\begin{lemma}\label{lem:nonexpansive}
The maps $H$ and $H_W$ are order-preserving and $\ell_1$-nonexpansive.
\end{lemma}
\begin{proof}
Order preservation follows from local monotonicity. For any one local
rule $\rho\in\{r_v,r_v^W\}$ and any two inputs $\xi,\eta\ge0$, the vector
$\rho(\xi)-\rho(\eta)$ has entries all of the same sign, and their sum is
$\xi-\eta$. Hence $\norm{\rho(\xi)-\rho(\eta)}=|\xi-\eta|$. The
triangle inequality and the fact that projection cannot increase the norm give
$\norm{H(z)-H(z')}\le\sum_{v\in V}|z_v-z_v'|=\norm{z-z'}$.
The same proof applies to $H_W$.
\end{proof}

For a rational $q\in(0,1)$, the exact and hybrid discounted maps are
$q$-contractions on the complete space $\mathbb R_{\ge0}^V$. Thus the
following equations have unique solutions:
\begin{equation}\label{eq:roots}
 x=\e_o+qH(x),\qquad y=\e_o+qH_W(y).
\end{equation}
We use a variant of the discounted formulation of~\cite{haslebacher2025},
keeping the source injection $\e_o$ undiscounted and introducing the hybrid
maps defined above. We compute $x$ by solving a sequence of hybrid equations.
The next subsection establishes the discount required for recovering
the walk; the algorithms of Section~\ref{sec:algorithm} apply to any
rational $q\in(0,1)$.

\subsection{Recovering integer departure counts}\label{sec:recovery}

For any input network define the phase product $P$, configuration bound
$M$, and discount denominator $Q$ by
\begin{equation}\label{eq:discount}
 P=\prod_{v\in V}d_v,\qquad M=nP,\qquad Q=4M^2,
 \qquad q=1-\frac1Q.
\end{equation}
The integers $P,M,Q$ have short binary representations. Suppose for now
that the network is stopping. Its walk must terminate:
if it were infinite, the nonempty set of vertices visited infinitely
often would be closed under every listed outgoing arc, since each entry
of the list at such a vertex is used infinitely often. This is impossible
because each internal vertex has a path to a terminal. Let $T$ denote the
number of moves before absorption. A pre-departure configuration consists
of the current internal vertex and all local phases, so there are at most
$M$ configurations. None can repeat in a terminating walk, and therefore
$T\le M$.

Let $u\in\mathbb Z_{\ge0}^V$ be the vector of actual departure counts, so
$\norm{u}=T$. Conservation at the end of the walk gives
$u=\e_o+H(u)$. The discounted map preserves the coordinatewise interval
$[0,u]$, since $\e_o+qH(u)\le u$. Its fixed point therefore satisfies
$0\le x\le u$.

\begin{lemma}\label{lem:rounding}
For the discount in~\eqref{eq:discount}, on a stopping network the exact
departure counts are $u_v=\lceil x_v\rceil$ for every $v\in V$.
\end{lemma}
\begin{proof}
For $k\ge0$, let $U^{(k)}$ count departures in the first $\min\{k,T\}$
moves, and let $p_k$ be the position after these moves. Conservation
along this prefix gives
\[
 \e_o+H(U^{(k)})-U^{(k)}=
 \begin{cases}
  \e_{p_k}, & k<T,\\
  0, & k\ge T.
 \end{cases}
\]
For $k<T$, exactly one departure occurs next, from $p_k$, so
\[
 U^{(k+1)}=U^{(k)}+\e_{p_k}=\e_o+H(U^{(k)}).
\]
For $k\ge T$, the counts remain equal to $u$ and satisfy the same
recurrence. Thus $U^{(0)}=0$ and $U^{(T)}=u$.
Compare these counts with the discounted iterates $X^{(0)}=0$ and
$X^{(k+1)}=\e_o+qH(X^{(k)})$. Writing
$E_k=\norm{U^{(k)}-X^{(k)}}$ and using Lemma~\ref{lem:nonexpansive}, we have
\begin{align*}
 E_{k+1}
 &\le q\norm{H(U^{(k)})-H(X^{(k)})}+(1-q)\norm{H(U^{(k)})}\\
 &\le qE_k+(1-q)k,
\end{align*}
since $\norm{H(U^{(k)})}\le\norm{U^{(k)}}\le k$. It follows from $E_0=0$
that $E_T\le(1-q)T(T-1)/2$. The discounted iterates increase from zero
and are bounded above by $x$, so $X^{(T)}\le x\le u$. Consequently,
\[
 0\le\norm{u-x}\le E_T\le\frac{T(T-1)}{8M^2}<\frac18.
\]
As each $u_v$ is integral, $u_v-1<x_v\le u_v$, proving the claim.
\end{proof}

Given $u$, the reached terminal is the unique terminal coordinate with
value one in $\sum_{v\in V}r_v(u_v)$; all other terminal coordinates are
zero. The iteration argument bounds the rounding error. Section~\ref{sec:algorithm} computes $x$ directly by solving hybrid systems with the discount in~\eqref{eq:discount}.

\section{Computing the discounted fixed point}\label{sec:algorithm}

\subsection{Period refinement}\label{sec:refinement}

Fix $q\in(0,1)$, and let $x,y$ be the exact and hybrid solutions of
\eqref{eq:roots}. In this subsection $x$ is fixed; only the sets $W_v$
and their corresponding hybrid solution $y$ change. The following lemma
is the main ingredient of our algorithm.

\begin{lemma}[A missing period]\label{lem:missing}
If $x\ne y$, there exist $v\in V$ and $k\notin W_v$ such that
\[
 x_v,y_v\in(kd_v,(k+1)d_v).
\]
\end{lemma}
\begin{proof}
Suppose first that the full local outflows $r_v(x_v)$ and $r_v^W(y_v)$
are coordinatewise comparable at every vertex (one is at most the other
in every coordinate). Their coordinate sums are $x_v$ and $y_v$, so in
this case their $\ell_1$ distance equals $|x_v-y_v|$. By
\eqref{eq:roots}, the triangle inequality, and nonexpansiveness of the
projection,
\[
 \norm{x-y}=q\norm{H(x)-H_W(y)}
 \le q\sum_{v\in V}\norm{r_v(x_v)-r_v^W(y_v)}
 =q\norm{x-y},
\]
contradicting $q<1$ and $x\ne y$. Hence some vertex has incomparable
local outflows. No period boundary can lie between the two inputs at
this vertex, including either endpoint. Indeed, if
$x_v\le jd_v\le y_v$, monotonicity and the common boundary
values in~\eqref{eq:boundary} give
\[
 r_v(x_v)\le r_v(jd_v)=jm_v=r_v^W(jd_v)\le r_v^W(y_v).
\]
If $y_v\le jd_v\le x_v$, the same argument gives
\[
 r_v^W(y_v)\le r_v^W(jd_v)=jm_v=r_v(jd_v)\le r_v(x_v).
\]
Either chain contradicts incomparability. The two inputs therefore lie
strictly inside the same period.
This period cannot be in $W_v$, because within a restored period both
outflows are evaluated under the same monotone rule $r_v$.
\end{proof}

Algorithm~\ref{alg:refine} starts with $W_v=\varnothing$ at every vertex,
solves the hybrid system, and checks the resulting point against the
exact equation. If the check fails, it restores the period containing
the current hybrid value at every vertex. By uniqueness, the exact rational equality test succeeds precisely when $y=x$. The periods to restore are determined by the current hybrid solution alone.

\begin{algorithm}[t]
\caption{Period refinement, $\Call{Refine}{q}$}\label{alg:refine}
\begin{algorithmic}
\Require A periodic routing network and rational $q\in(0,1)$
\Ensure The exact discounted fixed point $x$
\State Set $W_v\gets\varnothing$ for every $v\in V$
\Loop
  \State Construct the ordered affine-piece representation of each local map $\pi r_v^W$
  \State $y\gets\Call{SolveHybrid}{q,W}$
  \If{$y=\e_o+qH(y)$}
    \State \Return $y$
  \EndIf
  \ForAll{$v\in V$}
    \State $W_v\gets W_v\cup\{\lfloor y_v/d_v\rfloor\}$
  \EndFor
\EndLoop
\end{algorithmic}
\end{algorithm}

\begin{lemma}\label{lem:rounds}
Algorithm~\ref{alg:refine} makes at most $n$ refinements and at most
$n+1$ calls to \textnormal{\textsc{SolveHybrid}}.
\end{lemma}
\begin{proof}
For the fixed but unknown solution $x$, consider the potential
\[
 \Phi(W)=\bigl|\{v\in V:x_v\in(kd_v,(k+1)d_v)
                          \text{ for some }k\in W_v\}\bigr|.
\]
Each vertex has at most one open period containing its true coordinate.
If $x_v$ is a period boundary, $v$ cannot witness Lemma~\ref{lem:missing}.
Restored periods are never removed, so $\Phi$ is nondecreasing. After
each failed test, Lemma~\ref{lem:missing} supplies a vertex whose true
value lies in the same unrestored period as its hybrid value. Restoring
$\lfloor y_v/d_v\rfloor$ at every vertex includes that period and increases
$\Phi$ by at least one. Since $0\le\Phi(W)\le n$, at most $n$ tests can
fail. The next successful test returns $x$.
\end{proof}

\subsection{Monotone continuation for a hybrid system}\label{sec:continuation}

Fix a collection $W$ of restored periods. The function $r_v^W$ has a
finite representation by affine pieces: subdivide each restored period
into unit intervals, and use one averaged affine piece on each intervening
unrestored interval. Merge duplicate endpoints and leave a final unbounded
piece. Let $K$ denote the number of distinct positive finite endpoints,
counted separately at each vertex. We allow redundant endpoints between identical affine pieces. The solver follows a single path of fixed points. Define $y(\alpha)$ by
\begin{equation}\label{eq:path}
 y(\alpha)=\alpha\e_o+qH_W(y(\alpha)),\qquad 0\le\alpha\le1.
\end{equation}
The path starts at $y(0)=0$ and ends at the desired hybrid solution
$y(1)$ from~\eqref{eq:roots}.

\begin{lemma}\label{lem:continuation}
The path $y(\alpha)$ is continuous and coordinatewise nondecreasing.
On a product of local affine pieces, write $H_W(z)=Az+c$. The matrix
$A$ is nonnegative and its column sums are at most one. On the portion of
the path in this product, we have
\begin{equation}\label{eq:affine}
 y(\alpha)=a+\alpha b,\qquad
 G=(I-qA)^{-1},\quad a=qGc,\quad b=G\e_o\ge0.
\end{equation}
Given the ordered local affine-piece representations with $K$ represented
breakpoints, the point $y(1)$ can be computed exactly by processing at most
$K$ nonterminal breakpoint events and solving at most $K+1$ linear systems
of dimension $n$ (each with two right-hand sides). This uses
$O(n^3(K+1))$ exact rational arithmetic operations and comparisons.
\end{lemma}
\begin{proof}
If $0\le\alpha\le\alpha'\le1$, the fixed-point iterations from zero
for the two injections are coordinatewise ordered by monotonicity of
$H_W$. Taking their limits gives $y(\alpha)\le y(\alpha')$.
Nonexpansiveness and~\eqref{eq:path} also yield
\[
 (1-q)\norm{y(\alpha')-y(\alpha)}\le |\alpha'-\alpha|,
\]
proving continuity. Every local slope is nonnegative and has coordinate
sum one before projection, giving the column bound on $A$. Thus the Neumann series $G=\sum_{j\ge0}(qA)^j$ converges in the
induced matrix $\ell_1$ norm and gives a nonnegative inverse. Solving the
linear system gives~\eqref{eq:affine}. We now prove the algorithmic part
of the lemma. Start at the known fixed point $y(0)=0$.
At the current parameter $\alpha<1$, choose the local piece containing
$y_v(\alpha)$, taking the piece to its right whenever this value is an
endpoint. Denote the right endpoint of the selected piece at $v$ by
$\beta_v$, possibly $+\infty$. Every finite $\beta_v$ is strictly larger
than $y_v(\alpha)$. The affine solution in~\eqref{eq:affine} agrees with
the current point because the local rules are continuous at their
endpoints. Since $b\ge0$, this solution remains within the chosen
product of pieces until the parameter reaches
\begin{equation}\label{eq:event}
 \alpha'=\min\left(\{1\}\cup
   \left\{\alpha+\frac{\beta_v-y_v(\alpha)}{b_v}:
          v\in V,\ b_v>0,\ \beta_v<+\infty\right\}\right).
\end{equation}
Coordinates with $b_v=0$ remain constant until the next event.
All candidates exceed $\alpha$, so $\alpha'>\alpha$. On the entire
interval $[\alpha,\alpha']$, the affine candidate satisfies~\eqref{eq:path}
and therefore coincides with the unique fixed point. Unless
$\alpha'=1$, at least one coordinate has reached its next breakpoint.
Let $J$ be the set of all coordinates reaching their next endpoints,
and put $y_*=y(\alpha')$. Advance the local pieces for all $v\in J$
before computing the next affine solution. Write $A',c'$ for the matrix
and intercept after these changes. For each $v\in J$, let $\Delta_v$
be the new local slope column minus the old one. Continuity at
$\beta_v$ means that the corresponding local intercept changes by
$-\beta_v\Delta_v$. Thus
\[
 A'=A+\sum_{v\in J}\Delta_v\e_v^{\mathsf T},\qquad
 c'=c-\sum_{v\in J}\beta_v\Delta_v.
\]
Since $(y_*)_v=\beta_v$ for every $v\in J$, we have
\[
 A'y_*+c'=Ay_*+c+
 \sum_{v\in J}\Delta_v\bigl((y_*)_v-\beta_v\bigr)=Ay_*+c.
\]
Consequently $y_*=\alpha'\e_o+q(A'y_*+c')$: the same point solves the
new affine system at the event parameter. Every finite right endpoint
of the newly selected pieces is strictly larger than the corresponding
current coordinate. Thus the same invariant holds at the start of the next step,
even if a coordinate has zero slope or several coordinates reach their
endpoints simultaneously. Each nonterminal event processes at least one
previously unprocessed endpoint, and monotonicity prevents returning to
any earlier piece. Since there are only $K$ represented breakpoints, the
procedure must reach $\alpha=1$ after at most $K$ nonterminal events.

At initialization and after each nonterminal
event, form the matrix and intercept of the selected pieces and solve
\[
 (I-qA)a=qc,\qquad (I-qA)b=\e_o
\]
by ordinary exact Gaussian elimination with two right-hand sides. Each
such solve, including the assembly of $A,c$ and the scan for the next
event, uses $O(n^3)$ arithmetic operations and comparisons. There are at
most $K+1$ solves.
\end{proof}

Rank-one updates reduce the arithmetic cost of this solver.

\subsection{Rank-one updates}\label{sec:update}

When the local piece at vertex $v$ changes at endpoint $\beta$, only the
$v$th column of $A$ changes. Let $\Delta\in\mathbb R^V$ be the new slope
column minus the old one. Continuity at $\beta$ also determines the change
in the intercept, giving
\begin{equation}\label{eq:column}
 A'=A+\Delta\e_v^{\mathsf T},\qquad c'=c-\beta\Delta.
\end{equation}
The Sherman--Morrison identity~\cite{sherman1950} gives an update for the inverse and both affine coefficients.

\begin{lemma}\label{lem:update}
Let $G,a,b$ be as in~\eqref{eq:affine}, let $g=G\Delta$, and set
$\delta=1-qg_v$. Then $\delta\ne0$, and the coefficients associated with
$A',c'$ are
\begin{align}
 G'&=G+\frac{q}{\delta}g(\e_v^{\mathsf T}G),\label{eq:inverse}\\
 a'&=a+\frac{q(a_v-\beta)}{\delta}g,
 & b'&=b+\frac{qb_v}{\delta}g.\label{eq:coefficients}
\end{align}
All three updates together take $O(n^2)$ exact rational arithmetic
operations.
\end{lemma}
\begin{proof}
Both $I-qA$ and $I-qA'$ are invertible by the column bound of
Lemma~\ref{lem:continuation}. Since $(I-qA)g=\Delta$, we have
$(I-qA')g=(1-qg_v)\Delta=\delta\Delta$. If $\Delta\ne0$ then $g\ne0$, and
invertibility rules out $\delta=0$; if $\Delta=0$ then $\delta=1$.
Multiplying the right-hand side of~\eqref{eq:inverse} by $I-qA'$ gives
the identity, proving the inverse formula. This also gives
$G'\Delta=g/\delta$. Substitution into $a'=qG'c'$ and $b'=G'\e_o$, using
\eqref{eq:column}, proves~\eqref{eq:coefficients}. Computing $g$ and the
outer product costs $O(n^2)$ operations; the coefficient updates take
$O(n)$ operations.
\end{proof}

If several coordinates reach their next endpoints at the same parameter,
record the entire set before making any update, and process its columns
in an arbitrary order. Each intermediate matrix still consists of valid
local slope columns and thus has column sums at most one. Moreover, at
such an event $a_v+\alpha b_v=\beta$, and~\eqref{eq:coefficients} shows
that $a'+\alpha b'=a+\alpha b$. Hence these updates leave the current
point unchanged. After all tied columns have been processed, the selected
pieces have strictly larger right endpoints at every changed coordinate,
restoring the invariant used in~\eqref{eq:event}.

\begin{algorithm}[t]
\caption{Hybrid fixed point, $\Call{SolveHybrid}{q,W}$}\label{alg:hybrid}
\begin{algorithmic}
\Require A discount $q\in(0,1)$ and restored period sets $W$, with their ordered local affine-piece representations
\Ensure The hybrid fixed point $y=\e_o+qH_W(y)$
\State $\alpha\gets0$, $y\gets0$; select the local pieces to the right of zero
\State Form $A,c$ and next endpoints $\beta_v$; compute $G,a,b$ by~\eqref{eq:affine}
\While{$\alpha<1$}
  \State Compute $\alpha'$ by~\eqref{eq:event} using the current $\alpha,y,b,\beta_v$
  \State $\alpha\gets\alpha'$, $y\gets a+\alpha b$
  \If{$\alpha=1$}
    \State \Return $y$
  \EndIf
  \State $J\gets\{v\in V:y_v=\beta_v<+\infty\}$
  \ForAll{$v\in J$}
    \State $\beta\gets\beta_v$; let $\Delta$ be the next slope column minus the current one
    \State $g\gets G\Delta$, $\delta\gets1-qg_v$
    \State Update $G,a,b$ simultaneously by~\eqref{eq:inverse}--\eqref{eq:coefficients}
    \State Update $A,c$ by~\eqref{eq:column}; advance the local piece and $\beta_v$
  \EndFor
\EndWhile
\end{algorithmic}
\end{algorithm}

\begin{lemma}\label{lem:hybrid-cost}
Given the ordered local affine-piece representations of a hybrid system
with $K$ represented breakpoints,
Algorithm~\ref{alg:hybrid} computes its fixed point in
$O(n^3+n^2K)$ exact rational arithmetic operations and comparisons.
\end{lemma}
\begin{proof}
The continuation argument shows that the parameter strictly increases,
that the current point is always the exact solution at its parameter,
and that no local endpoint is processed more than once. The initial
inverse costs $O(n^3)$ operations, and Lemma~\ref{lem:update} gives
$O(n^2)$ per changed column. There are at most $K$ such changes; finding
the next event and collecting tied coordinates cost $O(n)$ per event.
This proves the bound and termination.
\end{proof}

\subsection{Arithmetic and bit complexity}\label{sec:complexity}

We now take the discount in~\eqref{eq:discount}. By Lemma~\ref{lem:rounds},
each vertex restores at most $n$ periods and there are at most $n+1$
hybrid solves. A restored period at vertex $v$ introduces at most $d_v+1$
breakpoints. Since $D\ge n$, the number of represented breakpoints in
any hybrid solve is at most
\begin{equation}\label{eq:budget}
 K\le n\sum_{v\in V}(d_v+1)=n(D+n)=O(nD).
\end{equation}
Thus Lemma~\ref{lem:hybrid-cost} bounds the total arithmetic cost of the
solves by $O(n(n^3+n^2\cdot nD))=O(n^4D)$. The representations constructed
before each call can be built by sorting restored indices and reading one copy of the list
for each restored period, with a single averaged piece for each intervening interval. To evaluate an exact rule at a rational $z\ge0$,
write $z=kd_v+j+\theta$, where $k=\lfloor z/d_v\rfloor$,
$j=\lfloor z-kd_v\rfloor\in\{0,\ldots,d_v-1\}$, and $0\le\theta<1$.
Then
\begin{equation}\label{eq:evaluate}
 r_v(z)=km_v+r_v(j)+\theta\e_{\sigma_v(j)}.
\end{equation}
The term $r_v(j)$ requires only a prefix of the explicit list at $v$.
Store multiplicities and outflows by successor, using the terminal coordinates appearing in the lists.
The integer parts may be computed by exact division (or binary search
using the size bounds below). Constructing pieces, testing the original
fixed-point equation, and restoring periods therefore require only
polynomial additional bit time. Storing only internal coordinates,
forming the local affine data takes $O(n(K+D+n))$ operations per solve.
Sorting the at most $n$ restored indices per vertex takes $O(n^2\log(n+1))$
comparisons per solve, and computing integer parts by binary search takes
$O(n(D+\log(n+1)))$ comparisons per refinement round. These costs are
dominated by the total arithmetic bound. With direct linear-system solves as in Lemma~\ref{lem:continuation}, the total arithmetic cost is
\[
 O\bigl((n+1)n^3(n(D+n)+1)\bigr)=O(n^5D),
\]
which is $O(n^6)$ for binary switches. This bound also covers the representation and testing costs.

To verify polynomial bit complexity, note that $\log_2 P\le D$ and
$Q=4(nP)^2$, so all integers in~\eqref{eq:discount}, and the numerator
and denominator of $q$, have $O(D+\log(n+1))$ bits. All hybrid continuation
solutions satisfy $\norm{y(\alpha)}\le\alpha+q\norm{y(\alpha)}$, hence
$\norm{y(\alpha)}\le Q$. Each restored period is selected at such a
solution, so its endpoints are at most $Q+d_{\max}$, where
$d_{\max}=\max_{v\in V}d_v$. For an exact local unit segment $[p,p+1]$,
where $p$ is a nonnegative integer, the slope is
$\e_{\sigma_v(p\bmod d_v)}$ and the intercept equals
$r_v(p)-p\e_{\sigma_v(p\bmod d_v)}$. After projection the slope is either
a unit vector or zero, and the integral intercept has entries of
magnitude at most $p\le Q+d_{\max}$. For an averaged piece the intercept
is zero and the slope denominator divides $d_v$. Consequently the
internal affine representation $Az+c$ has an integral $c$ with
$|c_i|\le n(Q+d_{\max})$ for every coordinate $i$, and each entry of $PA$
is integral.

Define the integer matrix $L=QP(I-qA)=QP I-(Q-1)PA$. The equations
for the affine coefficients become
\begin{equation}\label{eq:integer}
 LG=QP I,\qquad La=(Q-1)Pc,\qquad Lb=QP\e_o.
\end{equation}
All entries of these systems have $O(D+\log(n+1))$ bits.
Hadamard's inequality and Cramer's rule bound the reduced numerators and
denominators of every entry of $G,a,b$ by $O(n(D+\log(n+1)))$ bits.
The event parameters have the same asymptotic bit bound, since a finite
candidate in~\eqref{eq:event} can also be written
$(\beta_v-a_v)/b_v$. So do the coordinates of $a+\alpha b$ at these
parameters.

The determinant bound also controls intermediate quantities during Gaussian elimination. Work with the integer matrix $L$ and the appropriate integer right-hand sides in~\eqref{eq:integer}. Use Gaussian elimination with
row pivoting, eliminating below each pivot without scaling its row.
Reduce all fractions after each operation. Write
$B=O(D+\log(n+1))$ for a bound on the bit length of each input entry.
After $r$ pivots, let $L_r$ denote the nonsingular leading $r$ by $r$
block of the row-permuted original coefficient matrix. Each remaining
entry (including an entry in a right-hand-side column) has the form
$s-fL_r^{-1}h$, where $f,h,s$ are the corresponding row, column, and
entry of the row-permuted original augmented matrix. The block
determinant identity gives
\[
 s-fL_r^{-1}h=
 \frac{\det\begin{pmatrix}L_r&h\\ f&s\end{pmatrix}}{\det L_r}.
\]
The identity follows by subtracting $fL_r^{-1}$ times the first $r$
rows of the block matrix from its last row. For $r=0$, the expression
is just the original entry, with denominator one. The identity describes
the Schur complement produced by elimination;
row exchanges among the unprocessed rows only permute these entries.
Nonsingularity of $L$ guarantees that a nonzero pivot can always be
chosen. Hadamard's inequality bounds both determinants by integers of
$O(n(B+\log(n+1)))$ bits. Thus all reduced entries stored in forward
elimination have polynomial bit length, as do the multipliers and
intermediate products used in a single elimination operation. In back
substitution the completed solution coordinates have the same bound by
Cramer's rule; the partial sums involve at most $n$ products of such
polynomial-bit numbers and therefore have polynomial bit length even
before reduction. There are at most $n+2$ right-hand sides if $G,a,b$
are computed together, so elimination and back substitution use
$O(n^3)$ arithmetic operations and comparisons. Thus both the initial solve in the rank-one version and each direct linear-system solve take polynomial bit time in the explicit input length.

At every rank-one update, the resulting $G,a,b$ are the coefficients of
the current system, to which~\eqref{eq:integer} applies independently of
the preceding updates. The intermediate expressions in one update consist
of sums of at most $n$ products and a fixed number of further arithmetic
operations on polynomial-bit rational numbers. Reducing fractions using
the Euclidean algorithm therefore ensures polynomial bit length throughout.
This also covers the exact comparisons and divisions used for finding and
testing events, locating restored periods, and evaluating~\eqref{eq:evaluate}.
Together with the arithmetic bounds, these estimates prove polynomial
bit complexity for both solvers.

\begin{proof}[Proof of Theorem~\ref{thm:main}]
On a stopping network, Algorithm~\ref{alg:refine} with the discount
in~\eqref{eq:discount}, using Algorithm~\ref{alg:hybrid}, returns the
unique exact discounted solution $x$ within the bounds just proved.
Lemma~\ref{lem:rounding} recovers $u$ by rounding, and the terminal can
be identified by evaluating the local outflows at $u$.

For a general reachability instance, make the designated destination
absorbing and mark every vertex from which it is reachable in the
underlying directed graph. If the origin is the destination or is
unmarked, the answer is immediate. Otherwise replace all unmarked
vertices, including any other terminals, by a single absorbing failure
terminal. In the resulting network every retained internal vertex has
a path to the destination, so the network is stopping. The walk agrees
with the original until it either reaches the destination or enters an
unmarked vertex, from which the destination cannot be reached. Hence the
reduction preserves the answer and does not increase the number of
internal vertices or the total list length. Apply the stopping algorithm
with the reduced network's parameters. Since $D=2n$ for binary switches
before reduction, the claimed $O(n^5)$ arithmetic bound follows.
\end{proof}

\begin{corollary}[A solver without inverse updates]\label{cor:elementary}
The reachability and exact departure-count conclusions of
Theorem~\ref{thm:main} also hold when each hybrid fixed point is computed by
ordinary exact linear-system solves at the continuation events. This
version uses $O(n^5D)$ exact rational arithmetic operations and
comparisons, or $O(n^6)$ for binary switches, and has polynomial bit
complexity.
\end{corollary}
\begin{proof}
Use Lemma~\ref{lem:continuation} to solve the hybrid systems in
Algorithm~\ref{alg:refine}. Lemma~\ref{lem:rounds} still gives at most
$n+1$ calls, each with $O(nD)$ represented breakpoints by~\eqref{eq:budget}.
The bounds for direct linear-system solves and their bit complexity were
proved above. Lemma~\ref{lem:rounding} and the stopping reduction from
Theorem~\ref{thm:main} complete the proof.
\end{proof}

\section{Random access to the walk}\label{sec:random}

Departure counts also allow us to recover the state of a walk at a given time, in both stopping and nonstopping networks. An initially absorbing origin stays fixed. Otherwise write $T$ for its
absorption time, with $T=+\infty$ if it never terminates, and take the
configuration at time $t$ to be the position and phases after
$\min\{t,T\}$ moves. Once a terminal is reached, the configuration is
frozen.

\begin{theorem}\label{thm:random}
Given a binary-encoded integer $t\ge0$, the configuration of a periodic
routing network at time $t$ can be computed in time polynomial in the
explicit input length and $\log(t+2)$. The algorithm makes
$O(n\log(t+2))$ calls to the stopping-network algorithm, each adding
$O(\log(t+2))$ binary switches before the stopping reduction.
\end{theorem}

\begin{proof}
For $v\in V$ and integer $j\ge1$, let $\tau_v(j)$ denote the time of the
$j$th visit to $v$ immediately before departure, or $+\infty$ if this visit
never occurs. To compute $\tau_v(j)$, attach a binary counter with
$\ell=\lceil\log_2(j+1)\rceil$ new switches $C_0,\ldots,C_{\ell-1}$,
ordered from least to most significant bit. For $i<\ell-1$, the list at
$C_i$ is $(v,C_{i+1})$; at $C_{\ell-1}$ it is $(v,s_\star)$, where
$s_\star$ is a fresh terminal. A zero bit sends the token to $v$ and
becomes one; a one bit becomes zero and carries to the next bit (or
to $s_\star$ at the last bit). Initialize the counter phases to the bits
of $2^\ell-j$. Each input increments the counter; the first $j-1$
inputs pass to $v$, and the $j$th input overflows to $s_\star$. Redirect
all \emph{original} successor-list entries pointing to $v$ to $C_0$,
leaving the new counter's pass edges to $v$ unchanged. If $v=o$, move
the origin to $C_0$. Removing the counter moves from the augmented walk
recovers the original walk up to the intercepted visit to $v$.

Apply the stopping reduction from the proof of Theorem~\ref{thm:main}
with target $s_\star$. If the origin is unmarked or the reduced walk
does not reach $s_\star$,
then $\tau_v(j)=+\infty$. Otherwise sum the departure counts at retained
original internal vertices, obtained by Lemma~\ref{lem:rounding} in the
reduced instance. Every original vertex from which the augmented walk departs before reaching
$s_\star$ is retained, since the rest of the walk gives a path to $s_\star$. The sum counts the original moves up to the
intercepted visit, before the $j$th departure from $v$. Thus it equals
$\tau_v(j)$; in particular $\tau_o(1)=0$.

For $t=0$ return the initial configuration. For $t\ge1$, compute, at
each $v$, the number of departures strictly before time $t$ by binary
search for the largest $j\in\{1,\ldots,t\}$ with $\tau_v(j)<t$, taking
this number to be zero when the set is empty. Denote it by $N_v(t)$.
The search is valid because the visit times increase with $j$ until
they become infinite. The phase at $v$ has advanced by $N_v(t)\bmod d_v$,
relative to its initial phase. The position is the unique unit entry of
\begin{equation}\label{eq:position}
 \e_o+\sum_{v\in V}r_v(N_v(t))-
       \sum_{v\in V}N_v(t)\e_v\in\mathbb R^{V\cup S}.
\end{equation}
Indeed, every move subtracts one unit from its tail and adds one to its
head. The identity continues to hold after absorption.

Each search uses $O(\log(t+2))$ queries, and each augmented network
contains at most $n+\ell$ internal vertices and has phase product at
most $P2^\ell$. If the reduced walk reaches $s_\star$, its length is at
most $(n+\ell)P2^\ell$, so its departure counts and the computed finite
visit time have polynomial bit length. Theorem~\ref{thm:main}, the binary
searches, and the outflow evaluations in~\eqref{eq:position} therefore
use polynomial bit time in the claimed parameters.
\end{proof}
\section*{AI Use Statement}
ChatGPT (OpenAI) was used during the preparation of this manuscript to assist with drafting and revising the exposition, reviewing and refining mathematical arguments, checking notation, and bibliographic research.


\begin{thebibliography}{99}
\small

\bibitem{dohrau2017}
J\'er\^ome Dohrau, Bernd G\"artner, Manuel Kohler, Ji\v{r}\'i Matou\v{s}ek,
and Emo Welzl.
\newblock ARRIVAL: A zero-player graph game in \(\mathsf{NP}\cap\mathsf{coNP}\).
\newblock In \emph{A Journey Through Discrete Mathematics},
pages 367--374. Springer, 2017.

\bibitem{holroyd2008}
Alexander E. Holroyd, Lionel Levine, Karola M\'esz\'aros, Yuval Peres, James Propp, and David B. Wilson.
\newblock Chip-firing and rotor-routing on directed graphs.
\newblock In \emph{In and Out of Equilibrium 2}, Progress in Probability 60, pages 331--364. Birkh\"auser, 2008.

\bibitem{fearnley2021}
John Fearnley, Martin Gairing, Matthias Mnich, and Rahul Savani.
\newblock Reachability switching games.
\newblock \emph{Logical Methods in Computer Science}, 17(2):10:1--10:29, 2021.

\bibitem{karthik2017}
C. S. Karthik.
\newblock Did the train reach its destination: The complexity of finding a witness.
\newblock \emph{Information Processing Letters}, 121:17--21, 2017.

\bibitem{gartner2018}
Bernd G\"artner, Thomas Dueholm Hansen, Pavel Hub\'a\v{c}ek, Karel Kr\'al, Hagar Mosaad, and Veronika Sl\'ivov\'a.
\newblock ARRIVAL: Next stop in CLS.
\newblock In \emph{45th International Colloquium on Automata, Languages, and Programming (ICALP 2018)}, LIPIcs 107, pages 60:1--60:13. Schloss Dagstuhl, 2018.

\bibitem{gartner2021}
Bernd G\"artner, Sebastian Haslebacher, and Hung P. Hoang.
\newblock A subexponential algorithm for ARRIVAL.
\newblock In \emph{48th International Colloquium on Automata, Languages,
and Programming (ICALP 2021)}, volume 198 of LIPIcs,
pages 69:1--69:14. Schloss Dagstuhl, 2021.

\bibitem{haslebacher2025}
Sebastian Haslebacher.
\newblock ARRIVAL: Recursive framework \& \(\ell_1\)-contraction.
\newblock In \emph{52nd International Colloquium on Automata, Languages,
and Programming (ICALP 2025)}, volume 334 of LIPIcs,
pages 95:1--95:17. Schloss Dagstuhl, 2025.

\bibitem{ghorbani2026}
Ebrahim Ghorbani, Raul Lopes, and Matthias Mnich.
\newblock ARRIVAL in cube-root subexponential time.
\newblock arXiv preprint arXiv:2610.28241, 2026.

\bibitem{auger2022}
David Auger, Pierre Coucheney, and Loric Duhaz\'e.
\newblock Polynomial time algorithm for ARRIVAL on tree-like multigraphs.
\newblock In \emph{47th International Symposium on Mathematical Foundations of Computer Science (MFCS 2022)}, LIPIcs 241, pages 12:1--12:14. Schloss Dagstuhl, 2022.

\bibitem{auger2023}
David Auger, Pierre Coucheney, Loric Duhaz\'e, and Kossi Roland Etse.
\newblock Generalized ARRIVAL problem for rotor walks in path multigraphs.
\newblock In \emph{Reachability Problems (RP 2023)}, LNCS 14235, pages 183--198. Springer, 2023.

\bibitem{ghorbani2025}
Ebrahim Ghorbani, Jonah Leander Hoff, and Matthias Mnich.
\newblock A quasi-polynomial time algorithm for Multi-Arrival on tree-like multigraphs.
\newblock In \emph{42nd International Symposium on Theoretical Aspects of Computer Science (STACS 2025)}, LIPIcs 327, pages 39:1--39:19. Schloss Dagstuhl, 2025.

\bibitem{ladders2026}
Bernd G\"artner, Sebastian Haslebacher, and Hung P. Hoang.
\newblock Sinks and ladders: ARRIVAL and SSG with two vertices per level.
\newblock In \emph{13th International Conference on Fun with Algorithms (FUN 2026)}, LIPIcs 366, pages 19:1--19:16. Schloss Dagstuhl, 2026.

\bibitem{chen2024}
Xi Chen, Yuhao Li, and Mihalis Yannakakis.
\newblock Computing a fixed point of contraction maps in polynomial queries.
\newblock In \emph{56th Annual ACM Symposium on Theory of Computing (STOC 2024)}, pages 1364--1373. ACM, 2024.

\bibitem{haslebacher2025queries}
Sebastian Haslebacher, Jonas Lill, Patrick Schnider, and Simon Weber.
\newblock Query-efficient fixpoints of $\ell_p$-contractions.
\newblock \emph{TheoretiCS}, 5, article 17, pages 1--36, 2026. Preliminary version in FOCS 2025, pages 2037--2053.

\bibitem{batziou2025}
Eleni Batziou, John Fearnley, Spencer Gordon, Ruta Mehta, and Rahul Savani.
\newblock Monotone contractions.
\newblock In \emph{57th Annual ACM Symposium on Theory of Computing (STOC 2025)}, pages 507--517. ACM, 2025.

\bibitem{sherman1950}
Jack Sherman and Winifred J. Morrison.
\newblock Adjustment of an inverse matrix corresponding to a change in one element of a given matrix.
\newblock \emph{Annals of Mathematical Statistics}, 21(1):124--127, 1950.

\end{thebibliography}
\end{document}